\documentclass[aps,prd,twocolumn,nofootinbib,longbibliography]{revtex4-2}
\providecommand{\araa}{ARA\&A}
\providecommand{\apj}{ApJ}
\providecommand{\apjl}{ApJL}

\providecommand{\aap}{A\&A}
\providecommand{\mnras}{MNRAS}
\providecommand{\nat}{Nature}

\providecommand{\prd}{Phys.~Rev.~D}
\providecommand{\prl}{Phys.~Rev.~Lett.}

\usepackage[utf8]{inputenc}
\usepackage[T1]{fontenc}
\usepackage{amsmath,amssymb,amsfonts,mathtools}
\usepackage{bm}
\usepackage{graphicx}
\usepackage{hyperref}
\usepackage{booktabs} % Added for professional table formatting

\hypersetup{hidelinks}

\begin{document}

\title{A Gravitational Wave Background from Intermediate Mass Black Holes in AGN Disks}

\author{Chiara M. F. Mingarelli}
\affiliation{Department of Physics, Yale University, New Haven, CT 06520, USA}
\email{chiara.mingarelli@yale.edu}

\date{November 24, 2026}
\begin{abstract}
\noindent Intermediate mass black holes (IMBHs), formed by stellar mass black hole mergers in active galactic nucleus (AGN) disks, are expected to inspiral into their central supermassive black holes (SMBHs). These intermediate mass ratio inspirals (IMRIs) generate a stochastic gravitational-wave background in the mHz--decihertz band. We present the first estimate of the AGN--IMRI background based on an IMBH mass budget anchored to the observed rate of stellar mass binary black hole mergers. Using the LIGO/Virgo/KAGRA (LVK) merger rate from version 5.0 of the Gravitational-Wave Transient Catalog, we report a benchmark characteristic strain of $h_c^{\mathrm{bench}}(3\,{\rm mHz})=1.35^{+0.14}_{-0.13}\times10^{-21}$ and upper limit of $h_c^{\mathrm{UL}}(3\,{\rm mHz})=1.73\times10^{-20}$. This benchmark amplitude is comparable to a stellar mass extreme mass ratio inspiral background, motivating joint modeling of these backgrounds in the LISA global fit. The Galactic white dwarf foreground complicates its measurement, however, annual modulation and source statistics offer potential ways to distinguish these signals. The high-frequency cutoff depends on SMBH mass and spin, with rapid spin allowing the background to extend into the decihertz band. Together with the LVK merger rate, a detection or upper limit would test how efficiently stellar mass black hole mergers in AGN disks build IMBHs that inspiral into SMBHs, while offering insights in SMBH demographics.
\end{abstract}

\maketitle
% Keep spare column space at the bottom, including around the wide equation.
%\makeatletter
%\raggedbottom
%\@booleantrue\raggedcolumn@sw
%\makeatother

\section{Introduction}
\label{sec:intro}

In active galactic nucleus (AGN) disks, stellar mass black holes can merge repeatedly to form intermediate-mass black holes (IMBHs), which may then be captured by the central supermassive black hole (SMBH)~\cite{McKernan2012,Tagawa2020}. Ground-based gravitational-wave (GW) observations have revealed massive binary black hole (BBH) systems whose components may include second-generation black holes~\cite{Abbott2023, LVK2025}. For the most massive systems, including GW190521~\cite{IMBH2020} and GW231123~\cite{GW231123}, hierarchical growth in AGN disks is one proposed formation channel~\cite{FordMcKernan2025,Delfavero2025}. However, IMBH formation need not lead to capture at all: migration can stall, allowing merger remnants to remain in the disk or galactic nucleus~\cite{McKernan2012,Bellovary2016}.

\begin{figure*}[t]
    \centering
    \includegraphics[width=\linewidth,trim=0 240bp 0 140bp,clip]{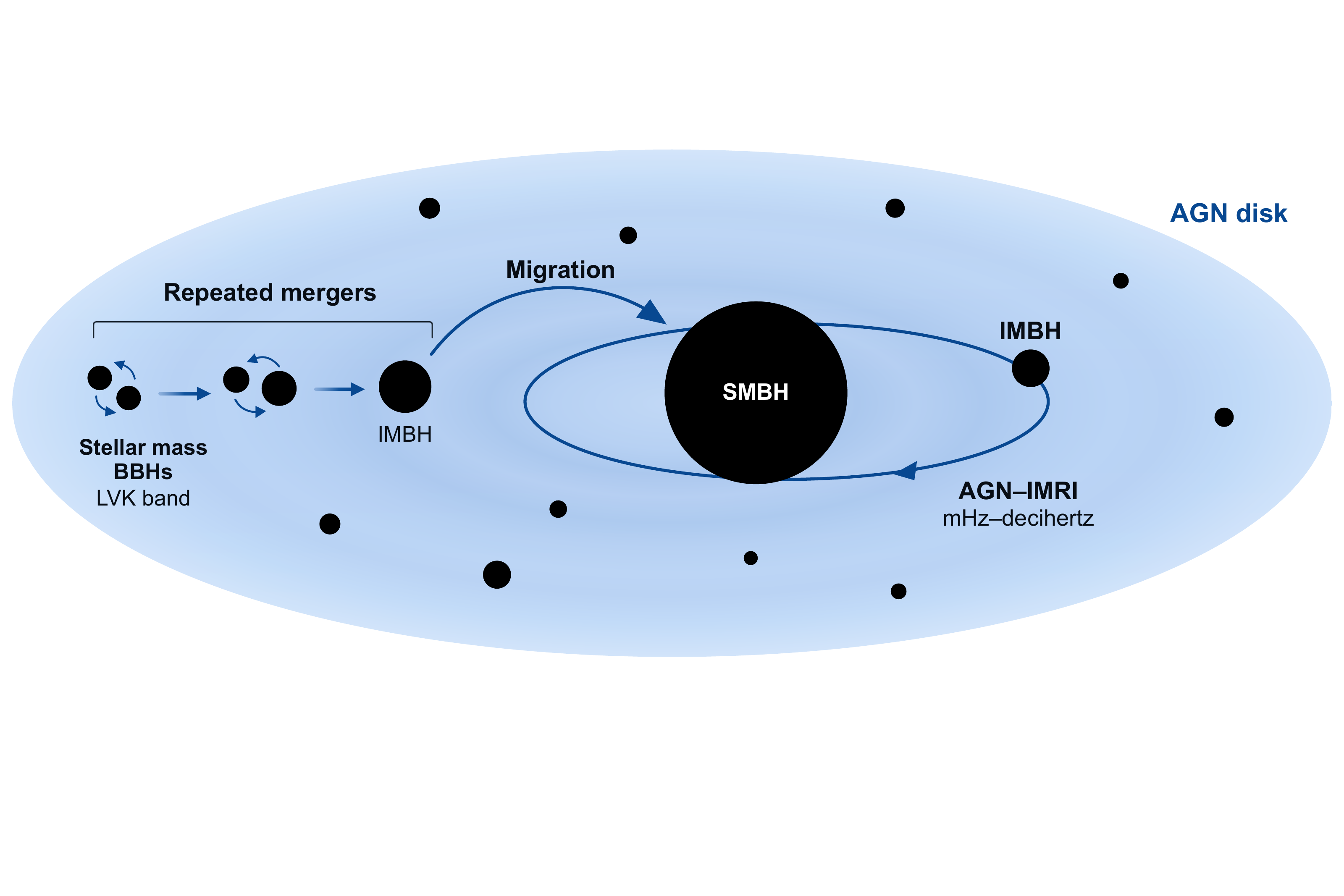}
    \caption{Proposed AGN--IMRI channel. Repeated stellar mass BBH mergers in an AGN disk can form IMBHs, which may migrate inward and inspiral into the central SMBH. The incoherent superposition of GW signals from IMBHs inspiralling into SMBHs generates the AGN--IMRI GWB in the mHz--decihertz band. We use the LVK merger rate and proposed AGN-channel efficiencies to estimate its amplitude and spectrum. Modeling the AGN--IMRI signal alongside stellar mass EMRIs and other astrophysical backgrounds can help identify and characterize the populations contributing to future GW observations. Its spectral features may help distinguish this component, while its high frequency cutoff could reveal the demographics and spins of the least massive SMBHs contributing to the signal. Figure inspired by Fig.~3 of Ref.~\cite{Tagawa2020}.}
    \label{fig:agn-imri-channel}
   \vspace{-1 \baselineskip}
\end{figure*}

If IMBHs do spiral into their central SMBHs, they produce the ``heavy''
intermediate mass ratio inspirals (IMRIs) considered here. In a ``light''
IMRI, a stellar mass compact object spirals into an IMBH~\cite{Gair2004}.
The GW signals from heavy IMRIs in AGN disks combine to produce a
stochastic GW background (GWB) in the mHz--decihertz band
(Fig.~\ref{fig:agn-imri-channel}). We call this the AGN--IMRI GWB.
We estimate its amplitude and spectral shape and examine how it could
be distinguished from other astrophysical backgrounds.
LISA~\cite{AmaroSeoane2017} and proposed decihertz missions such as the
Big Bang Observer (BBO)~\cite{Harry2006} and the DECi-hertz Interferometer
Gravitational wave Observatory (DECIGO)~\cite{Kawamura2021} target
different parts of this band.

SMBHs grow by accreting gas~\cite{TucciVolonteri2017}, capturing compact objects, and merging with other SMBHs. For the capture channel, comparing the captured IMBH mass density with the local SMBH mass density~\cite{Liepold2024} lets us assess the channel's contribution to SMBH growth.

Sigl et al.~\cite{Sigl2007} estimated the GWB from compact objects inspiraling into SMBHs in AGN disks, using the AGN X-ray luminosity function to infer their inspiral rates. Here we use the LIGO/Virgo/KAGRA (LVK) merger rate from the Gravitational-Wave Transient Catalog (GWTC-5.0)~\cite{GWTC5pop} and assumed AGN-channel efficiencies to estimate the cumulative mass density of captured IMBHs. We apply the formalism developed in our companion paper~\cite{Mingarelli2026} to convert this mass density into a GWB, deriving an empirically anchored benchmark amplitude and upper limit without requiring detailed modeling of AGN disk microphysics.

A measurement or upper limit of the AGN--IMRI GWB amplitude would give (or constrain) the cumulative mass density of IMBHs captured by central SMBHs. Combined with the LVK merger rate, this would test how efficiently stellar mass BBH mergers in AGN disks supply IMBH mass to central SMBHs.
Since inspirals into lower-mass SMBHs can reach higher GW frequencies
at the innermost stable circular orbit (ISCO), the high-frequency cutoff
could offer a new way to probe the least massive SMBHs that capture IMBHs.

At $3\,{\rm mHz}$ in Fig.~\ref{fig:imbh-smbh-emri}, our benchmark AGN--IMRI strain is comparable to a fiducial prediction by Bonetti and Sesana~\cite{BonettiSesana2020} (from hereon BS20) for the background from unresolved stellar mass extreme mass ratio inspirals (EMRIs). We show that annual modulation allows the Galactic double white dwarf (DWD) foreground~\cite{CornishRobson2017} to be separated from an isotropic background in our model. The harder problem for LISA's global fit~\cite{LittenbergCornish2023,AdamsCornish2014,Boileau2021Galactic} is distinguishing the AGN--IMRI and stellar mass EMRI backgrounds, since both are approximately isotropic. Differences in their spectra or in how often individual inspirals overlap may help distinguish them.

The rest of the paper is laid out as follows. In Sec.~\ref{sec:IMBH-SMBHcapture}, we derive the AGN--IMRI
strain from the captured IMBH mass density. We also consider cluster
infall as an alternative way to deliver IMBHs to SMBHs.
To interpret this signal in LISA data, we examine how annual modulation separates the
Galactic DWD foreground from an isotropic background in Sec.~\ref{sec:fisher_wd_imbh}.
We then calculate the benchmark amplitude and upper limit and compare
our spectra with stellar mass EMRI models in Sec.~\ref{sec:results}.
In Sec.~\ref{sec:discussion}, we discuss what the amplitude and cutoff
could reveal about SMBH growth and the least massive SMBHs that capture IMBHs, and how the
two inspiral populations might be distinguished. We conclude in Sec.~\ref{sec:conclusion}.
The \hyperref[app:capture-history]{Appendix} explores how the benchmark changes
if the capture rate evolves with redshift. We use $G=c=1$.
\par%

% Keep the Section II heading and opening sentence together.
%\begin{samepage}
\section{The AGN--IMRI Background}
\label{sec:IMBH-SMBHcapture}
\label{sec:alternative_channels}% former section now merged with the channel model

To estimate the amplitude of this AGN--IMRI GWB, we first express the characteristic strain in terms of the GW energy budget~\cite{Phinney2001, Mingarelli2026}:
%\par\end{samepage}
{
\begin{align}
\rho_c\Omega_{\rm gw}(f)
&=\frac{{\rm d}\rho_{\rm gw}}{{\rm d}\ln f}
=\frac{\pi f^2}{4}h_c^2(f)=\epsilon_{\rm gw}\rho_{\rm src}\mathcal H(f).
\label{eq:energy_distribution}
\end{align}
}
Here $\Omega_{\rm gw}(f)$ is the GW energy density per logarithmic frequency interval relative to the critical density $\rho_c=3H_0^2/(8\pi)$, $H_0$ is the Hubble constant, $\rho_{\rm gw}$ is the GW energy density, and $h_c(f)$ is the characteristic strain.
A fraction $\epsilon_{\rm gw}$ of the captured IMBH mass is radiated as GWs.
Following Ref.~\cite{Mingarelli2026}, we define the dimensionless function
\begin{equation}
\mathcal H(f) \equiv
\frac{1}{\epsilon_{\rm gw}\rho_{\rm src}}
\frac{{\rm d}\rho_{\rm gw}}{{\rm d}\ln f},
\end{equation}
where $\epsilon_{\rm gw}\rho_{\rm src}$ is the total GW energy emitted by all captures per comoving volume.
The spectral function satisfies $\int_0^\infty\mathcal H(f)\,{\rm d}\ln f\leq1$.
We denote the source frame frequency by $f_r=(1+z)f$, where $f$ is the observed frequency and $z$ is the source redshift.

\begin{table*}[ht!]
\centering
\caption{Benchmark values and explored ranges for the AGN--IMRI model. The BBH rate interval is a 90\% credible interval; the SMBH-density uncertainty is from \citet{Liepold2024}. The remaining ranges are model choices. The lookback intervals $T_{\rm eff}=1$, 10, and 13~Gyr correspond to starting redshifts $z_{\max}\simeq 0.07$, 1.66, and 5.84, respectively. The upper-limit calculation holds $\epsilon_{\rm gw}=0.05$ and $f_{\max}=0.04$\,Hz fixed.}
\label{tab:parameters}
\begin{tabular}{lclp{3.6in}}
\toprule
Parameter & Benchmark & Range & Physical meaning \\
\midrule
$\epsilon_{\rm gw}$ & 0.05 & 0.01--0.30 & GW radiative efficiency. \\
$f_{\rm AGN}$ & 0.1 & 0.01--1.0 & Fraction of BBH merger events occurring in AGN disks. \\
$\kappa_{\rm IM}$ & 0.3 & 0.05--1.0 & Efficiency of IMBH retention and eventual capture by the central SMBH. \\
$R_{\rm BBH}^{\rm obs}$ & $30.4^{+6.5}_{-5.4}$ & $25.0$--$36.9$; $49.4$ (upper limit) & LVK BBH merger rate density [Gpc$^{-3}$ yr$^{-1}$], at $z=0.2$~\cite{GWTC5pop}. \\
$\langle M_{\rm BBH}\rangle$ & 60 & 20--150 & Mean total mass of AGN-hosted BBH mergers [$M_\odot$]. \\
$T_{\rm eff}$ & 10 & 1--13 & Lookback interval over which the source-frame merger rate is held constant [Gyr]. \\
$M_{\bullet}^{\rm min}$ & $10^5$ & $10^4$--$10^7$ & Minimum SMBH mass for IMBH captures [$M_\odot$] (sets $f_{\max}$). \\
$\rho_\bullet$ & $1.8^{+0.8}_{-0.5} \times 10^6$ & $(1.3$--$2.6) \times 10^6$ & Local SMBH mass density [$M_\odot\,{\rm Mpc}^{-3}$],~\citet{Liepold2024}\\
\bottomrule
\end{tabular}
\end{table*}

We now calculate the total comoving mass density in IMBHs that inspiral into SMBHs, $\rho_{\rm IMR}=\rho_{\rm src}$. We take the total BBH merger rate density inferred by LVK, $R_{\rm BBH}^{\rm obs}$, to be constant in the source frame over a lookback interval $T_{\rm eff}$. If a fraction $f_{\rm AGN}$ of these mergers occur in AGN disks, then $f_{\rm AGN}R_{\rm BBH}^{\rm obs}T_{\rm eff}$ is their number per comoving volume. These mergers process a mean total binary mass $\langle M_{\rm BBH}\rangle$ per event. A fraction $\kappa_{\rm IM}$ of the processed mass ultimately reaches the central SMBHs in IMBHs. Thus $\rho_{\rm IMR}$ is the product of the number of AGN disk mergers per comoving volume and the mean mass delivered per merger, $\kappa_{\rm IM}\langle M_{\rm BBH}\rangle$.
\begin{equation}
\rho_{\rm IMR}
=\kappa_{\rm IM}\,
f_{\rm AGN}\,R_{\rm BBH}^{\rm obs}\,
\langle M_{\rm BBH}\rangle\,
T_{\rm eff} \, .
\label{eq:captured_mass}
\end{equation}
We treat $f_{\rm AGN}$ and $\kappa_{\rm IM}$ as model inputs, with $\kappa_{\rm IM}$ accounting for remnants that never reach the central SMBH, including those lost through ejection or disk exit.
For a fixed AGN disk mass supply, $\kappa_{\rm IM}\leq 1$ is conservative in that it can only reduce the signal relative to complete capture. 
We discuss growth by gas accretion separately in Sec.~\ref{sec:discussion}. Table~\ref{tab:parameters} summarizes the observational inputs and model assumptions.

Cluster infall offers another way to deliver IMBHs to central SMBHs.
Examples include globular clusters sinking by dynamical friction~\cite{ArcaSeddaGualandris2018} and merging nuclear star clusters~\cite{Ogiya2020}. An IMBH can form in a cluster core through runaway stellar
collisions~\cite{PortegiesZwart2004} or hierarchical black hole mergers~\cite{Greene2020}.
In the ``dry'' channel, tidal disruption of the infalling globular cluster
leaves the IMBH near the galactic center~\cite{ArcaSeddaGualandris2018}.
If IMBHs are common in clusters, this route can supply them to galactic nuclei, where they may interact with an AGN disk~\cite{McKernan2012}.
Equation~\eqref{eq:captured_mass} counts only the AGN disk contribution. Captures of IMBHs from nuclear star clusters or globular clusters would contribute a separate term to the captured mass density, which we leave for future work.
\par%

The captured IMBH mass density in Eq.~\eqref{eq:captured_mass} gives a total emitted GW energy per comoving volume of $\epsilon_{\rm gw}\rho_{\rm IMR}$. At the reference frequency, Eq.~\eqref{eq:energy_distribution} then gives
{
\begin{equation}
h_c^{2}(f_{\rm ref}) =
\frac{4}{\pi f_{\rm ref}^2}\epsilon_{\rm gw}\rho_{\rm IMR}
\mathcal H(f_{\rm ref}),
\label{eq:AIMR_exact}
\end{equation}
where $f_{\rm ref}=3\,{\rm mHz}$ is the reference frequency.
We evaluate $\mathcal H(f)$ for a constant source-frame rate over $T_{\rm eff}$, using a circular inspiral spectrum with source-frame cutoff $f_{\max}=0.04\,\mathrm{Hz}$, as in the companion paper~\cite{Mingarelli2026}.
For the 10~Gyr capture history, $\mathcal H(3\,{\rm mHz})\simeq 0.10$, so about $1\%$ of the emitted GW energy lies between 2.85 and 3.15~mHz today.

}
To test whether the captures that produce this GWB could supply an appreciable share of the mass in SMBHs today, we define 
\begin{equation}
f_{\rm IMR}\equiv\rho_{\rm IMR}/\rho_\bullet \, ,
\label{eq:captured_mass_fraction}
\end{equation}
where $\rho_\bullet$ is the local SMBH mass density (see, e.g., Refs.~\cite{Liepold2024,LiepoldMa2026Erratum}).

We adopt a minimum SMBH mass of $M_\bullet^{\rm min}=10^5\,M_\odot$.
For a circular, equatorial, prograde inspiral, the frequency of the dominant GW harmonic at the ISCO is~\cite{Bardeen1972}
\begin{equation}
f_{\rm ISCO}(M_\bullet,a_\bullet)
= \frac{1}{\pi M_\bullet
  \bigl[\hat r_{\rm ISCO}(a_\bullet)^{3/2}+a_\bullet\bigr]},
\label{eq:fISCO}
\end{equation}
where $a_\bullet$ is the dimensionless SMBH spin and $\hat r_{\rm ISCO}$ is the ISCO radius in units of $M_\bullet$.
For a nonspinning SMBH, $\hat r_{\rm ISCO}=6$, so our minimum SMBH mass gives $f_{\max}=f_{\rm ISCO}(M_\bullet^{\rm min},0)\simeq 0.04\,{\rm Hz}$.
Following our companion paper~\cite{Mingarelli2026}, we use this cutoff and $\epsilon_{\rm gw}=0.05$ for both the benchmark and upper-limit calculations.
In Sec.~\ref{sec:results}, we compare the GWB spectra for nonspinning SMBHs and SMBHs with prograde spin $a_\bullet=0.98$ (Fig.~\ref{fig:agn-imri-spin}).

\section{Separating the Galactic DWD Foreground}
\label{sec:fisher_wd_imbh}

A key challenge will be distinguishing the AGN--IMRI background from the Galactic DWD foreground and other extragalactic backgrounds (e.g., the stellar mass EMRI background in BS20).

%The Galactic white-dwarf foreground complicates measurement in the mHz band, and the predicted amplitude overlaps the range of stellar mass EMRI GWB models

We use a two-amplitude Fisher calculation to examine how LISA's annual modulation of the Galactic DWD foreground helps distinguish it from the AGN--IMRI background, despite their similar inspiral spectra.

Related calculations have jointly estimated Galactic foregrounds, isotropic backgrounds, and instrumental noise using Bayesian methods and Fisher analyses~\cite{AdamsCornish2014,Boileau2021Galactic}. Adams and Cornish~\cite{AdamsCornish2014} also considered backgrounds with identical spectra. Breivik et al.~\cite{BreivikMingarelliLarson2020} calculated the Galactic modulation harmonics and their measurability to constrain Galactic structure. We use the same modulation information, but hold the foreground shape and noise model fixed to isolate the uncertainty from fitting the two amplitudes, providing an analytic illustration of AGN--IMRI separation.

The extragalactic background can be approximated as isotropic, while the Galactic foreground is concentrated toward the Galactic plane. As LISA orbits the Sun, its antenna pattern rotates relative to the Galaxy, imposing an annual modulation on the measured Galactic power which is absent for an extragalactic background.

We decompose the real parts of auto- and cross-spectral estimates of the time-delay-interferometry channels into Fourier harmonics of the LISA orbital frequency $f_m = 1\,\text{yr}^{-1}$. We use independent real Fourier coefficients to express the spectral
power as its mean over one LISA orbit plus sine and cosine terms at
$k\neq0$. We call the constant term ($k=0$) the direct-current (DC)
component by analogy with a steady current.

The indices $I$ and $J$ label LISA's $A$, $E$, and $T$ channels,
each formed by combining time-delayed laser measurements between
spacecraft~\cite{BreivikMingarelliLarson2020}. For example, $AA$ and $EE$ are the power spectra of the
individual channels, while $AE$ is their cross spectrum.

In a frequency bin centered on $f$, the expected signal in the $k$-th harmonic for channel pair $IJ$ is:
\begin{equation}
\mu_k^{IJ}(f) = \left[A_{\rm iso}\,\mathcal R_{\rm iso}^{IJ}(f)\delta_{k0} + A_{\rm gal}\,\mathcal{M}_k^{IJ}(f)\right]S(f),
\label{eq:datamodel}
\end{equation}
where $A_{\rm iso}$ and $A_{\rm gal}$ are dimensionless power amplitudes, $S(f)$ is the common spectral shape with strain power spectral density units, $\mathcal R_{\rm iso}^{IJ}$ is the isotropic response, and $\mathcal{M}_k^{IJ}(f)$ is the modulation transfer function encoding the LISA response to the Galactic anisotropy. The Kronecker delta $\delta_{k0}$ reflects the fact that an isotropic background produces no modulation and contributes only to $k = 0$.

The modulation coefficients $\mathcal{M}_k^{IJ}$ depend on the Galactic geometry (primarily the $l = 2$ and $l = 4$ spherical harmonic moments of the DWD distribution) and the LISA orbit~\cite{UngarelliVecchio2001,Seto2004}.

We work with log-amplitudes $\boldsymbol{\theta} = (\ln A_{\rm iso}, \ln A_{\rm gal})$. The derivatives of the data model are:
\begin{align}
\frac{\partial \mu_k^{IJ}}{\partial \ln A_{\rm iso}} &= A_{\rm iso}\,\mathcal R_{\rm iso}^{IJ}(f)\delta_{k0}\,S(f), \\
\frac{\partial \mu_k^{IJ}}{\partial \ln A_{\rm gal}} &= A_{\rm gal}\,\mathcal{M}_k^{IJ}\,S(f).
\end{align}

Assuming Gaussian errors, approximating these coefficient estimates as uncorrelated, and holding their variances $\sigma_{k,IJ}^2(f)$ fixed at a fiducial model, the Fisher information matrix is:
\begin{equation}
\mathcal{F}_{ab} = \sum_f \sum_{IJ} \sum_k \frac{1}{\sigma_{k,IJ}^2(f)} \frac{\partial \mu_k^{IJ}}{\partial \theta_a} \frac{\partial \mu_k^{IJ}}{\partial \theta_b}.
\end{equation}
The variances include both instrumental noise and foreground fluctuations. For brevity, define the weights $w_k^{IJ}(f)=S^2(f)/\sigma_{k,IJ}^2(f)$.

Evaluating the sums over $k$, the $k = 0$ mode contributes to all elements while $k \neq 0$ modes contribute only to $\mathcal{F}_{22}$:
{
\begin{align}
\mathcal{F}_{11}
&= A_{\rm iso}^2 \sum_{f,IJ} w_0^{IJ}(\mathcal R_{\rm iso}^{IJ})^2
\equiv A_{\rm iso}^2 W, \\
\mathcal{F}_{12}
&=\! A_{\rm iso} A_{\rm gal} \sum_{\mathclap{f,IJ}} w_0^{IJ}\mathcal R_{\rm iso}^{IJ}\mathcal M_0^{IJ}
\!\equiv\! A_{\rm iso} A_{\rm gal} W\bar{\mathcal M}_0, \\
\mathcal{F}_{22}
&= A_{\rm gal}^2 \sum_{f,IJ,k} w_k^{IJ}(\mathcal M_k^{IJ})^2
\equiv A_{\rm gal}^2 W\bar{\mathcal M}_{\rm tot}^2.
\end{align}
}
where we have defined the total weight $W$, the mean DC response $\bar{\mathcal{M}}_0$, and the total modulation power $\bar{\mathcal{M}}_{\rm tot}^2$ (all averaged over frequencies and channel pairs). Here $\bar{\mathcal M}_0$ is the weighted projection of the Galactic DC response onto the isotropic response. We further decompose:
\begin{equation}
\bar{\mathcal{M}}_{\rm tot}^2 = \bar{\mathcal{M}}_0^2 + \bar{\mathcal{M}}_{\perp}^2,
\end{equation}
where the independent Galactic information is
{
\begin{align}
\bar{\mathcal M}_{\perp}^2
&=\frac{1}{W}\sum_{f,IJ}w_0^{IJ}
\left(\mathcal M_0^{IJ}-\bar{\mathcal M}_0\mathcal R_{\rm iso}^{IJ}\right)^2
\nonumber\\
&\quad+\frac{1}{W}\sum_{f,IJ}\sum_{k\ne0}w_k^{IJ}(\mathcal M_k^{IJ})^2.
\end{align}
}
The second term is the power in the modulation sidebands. The first accounts for differences between the Galactic and isotropic DC responses across frequencies or channel pairs. It vanishes when those DC responses are proportional, in which case $\bar{\mathcal M}_{\perp}^2$ reduces to the sideband power alone.

The Fisher matrix is therefore
\begin{equation}
\mathcal{F} = W \begin{pmatrix}
A_{\rm iso}^2 & A_{\rm iso} A_{\rm gal} \bar{\mathcal{M}}_0 \\[4pt]
A_{\rm iso} A_{\rm gal} \bar{\mathcal{M}}_0 & A_{\rm gal}^2 \bar{\mathcal{M}}_{\rm tot}^2
\end{pmatrix} \, ,
\label{eq:fishermatrix}
\end{equation}
and the determinant is:
\begin{equation}
\begin{split}
\det \mathcal{F}
&= W^2 A_{\rm iso}^2 A_{\rm gal}^2
\left(\bar{\mathcal{M}}_{\rm tot}^2-\bar{\mathcal{M}}_0^2\right)\\
&= W^2 A_{\rm iso}^2 A_{\rm gal}^2\,\bar{\mathcal{M}}_{\perp}^2.
\end{split}
\end{equation}
Inverting the Fisher matrix gives the marginalized uncertainty in $\ln A_{\rm iso}$,
\begin{equation}
\sigma_{\ln A_{\rm iso}} = \frac{1}{A_{\rm iso}\sqrt{W}} \frac{\bar{\mathcal{M}}_{\rm tot}}{\bar{\mathcal{M}}_{\perp}}.
\label{eq:sigma_Aiso}
\end{equation}

For our fixed Galactic template, the annual modulation constrains how
much of the mean power comes from the foreground. Fitting the Galactic
amplitude increases the uncertainty in the isotropic amplitude by a
factor of $\bar{\mathcal M}_{\rm tot}/\bar{\mathcal M}_{\perp}$
relative to the case in which the Galactic amplitude is known.

If the templates become proportional ($\bar{\mathcal{M}}_{\perp} \to 0$), the error diverges: $\sigma_{\ln A_{\rm iso}} \to \infty$. This is the statement that without modulation or distinct DC responses, the two components are perfectly degenerate in this fit. The correlation coefficient between $\ln A_{\rm iso}$ and $\ln A_{\rm gal}$ is:
\begin{equation}
r = -\frac{\mathcal{F}_{12}}{\sqrt{\mathcal{F}_{11}\mathcal{F}_{22}}} = -\frac{\bar{\mathcal{M}}_0}{\bar{\mathcal{M}}_{\rm tot}}.
\end{equation}
When $\bar{\mathcal{M}}_{\perp} \to 0$, we have $\bar{\mathcal{M}}_{\rm tot} \to |\bar{\mathcal{M}}_0|$ and $|r| \to 1$.

The modulation sidebands provide independent information about the Galactic foreground. With the same independent-coefficient approximation, their combined signal-to-noise ratio (SNR) is
\begin{equation}
\mathrm{SNR}_{\rm mod}^2
= A_{\rm gal}^2\sum_f\sum_{IJ}\sum_{k\ne0}
\frac{[\mathcal M_k^{IJ}S(f)]^2}{\sigma_{k,IJ}^2(f)}.
\label{eq:snr_mod}
\end{equation}
\begin{figure*}[ht!]
    \centering
    \includegraphics[width=\linewidth]{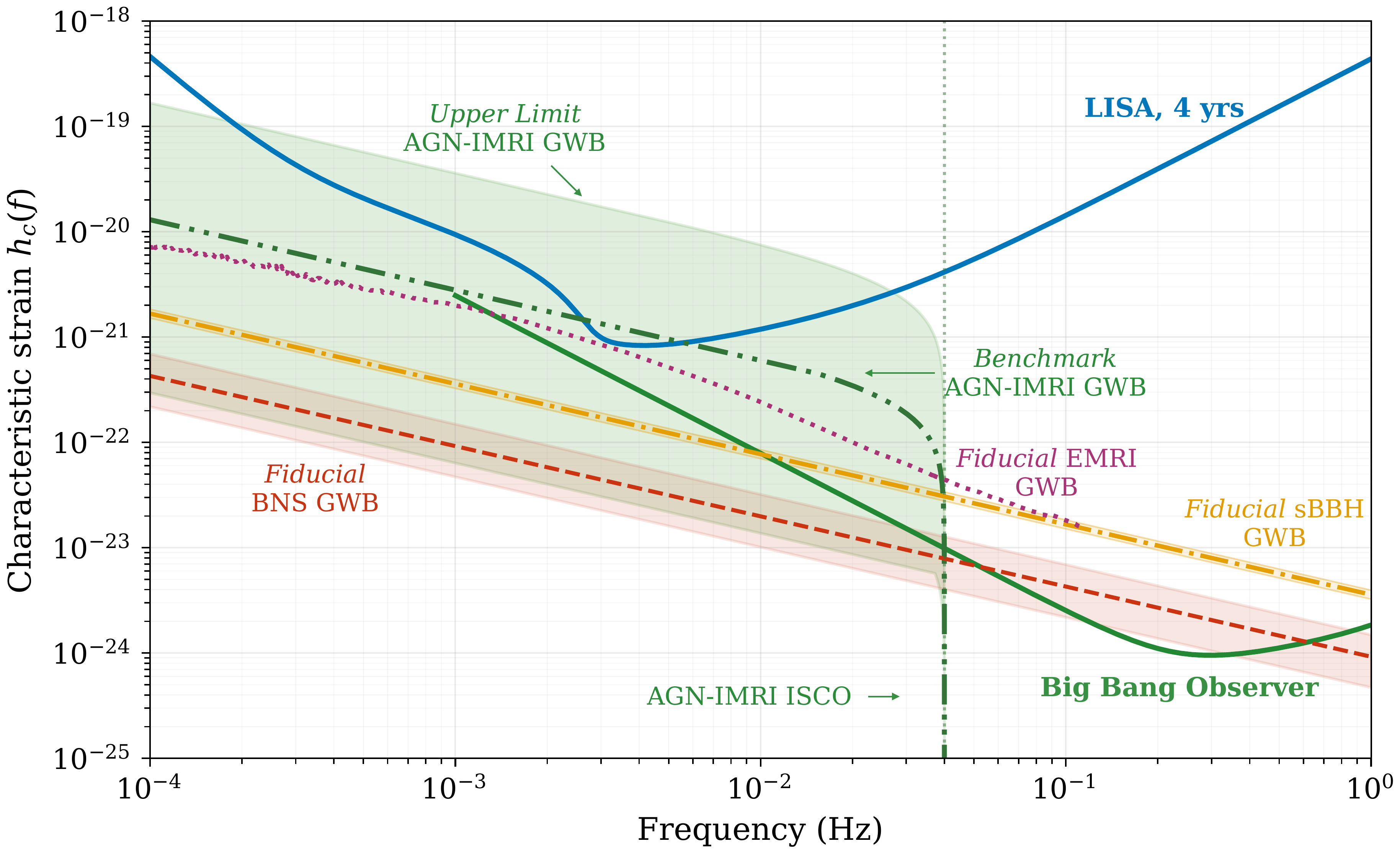}
           \vspace{-1 \baselineskip}
    \caption{Characteristic strain spectra of the AGN--IMRI background and other astrophysical backgrounds, compared with representative LISA~\cite{AmaroSeoane2017} and BBO sensitivities~\cite{Schmitz2021}. We obtain the green AGN--IMRI band by varying the five factors in
Eq.~\eqref{eq:captured_mass} over the ranges in Table~\ref{tab:parameters}.
We hold the GW efficiency fixed and set the source-frame cutoff to the
ISCO frequency of a nonspinning $10^5\,M_\odot$ SMBH. The dash--dot--dot curve
marks the benchmark and the upper edge marks the model upper limit. The stellar mass EMRI curve is the nonspinning M10 model, extracted from Fig.~3 of BS20. BS20 removes individually resolved sources with $\mathrm{SNR}>20$ over four years, whereas our AGN--IMRI curve includes all captures. The $0.04$\,Hz cutoff applies to our AGN--IMRI model; the BS20 EMRI background can extend to higher frequencies because it includes lower-mass SMBHs and higher harmonics from eccentric orbits. The stellar mass binary black hole (sBBH)~\cite{Zhu2011} and binary neutron star (BNS) benchmarks~\cite{Regimbau2011,LIGOScientific2018stochBNS,EbersoldRegimbau2025} use the formalism in our companion paper~\cite{Mingarelli2026}, before resolved-source subtraction; their faint bands propagate the GWTC-5.0 90\% merger-rate intervals with other inputs fixed~\cite{GWTC5pop}. The LISA sensitivity includes Galactic DWD confusion noise for a four-year observation. Figure~\ref{fig:agn-imri-spin} explores spin dependence.}
       \vspace{-1 \baselineskip}
    \label{fig:imbh-smbh-emri}
\end{figure*}

The estimator variance $\sigma_{k,IJ}^2(f)$ already accounts for
the observation time and bandwidth.

To assess whether LISA could measure the annual modulation used in our
fit, we use the Galactic DWD models of Breivik et al.~\cite{BreivikMingarelliLarson2020}.
For their full DWD population before source subtraction, they find SNRs
of $\mathcal{O}(10^2)$ for the strongest nonzero $AA$ and $EE$ harmonics over
four years in a $1\,{\rm mHz}$ bandwidth (see their Fig.~4).
Measuring these harmonics would constrain the Galactic contribution to
the mean power. Even a small uncertainty in that contribution could be
comparable to a weaker isotropic signal. See Eq.~\eqref{eq:sigma_Aiso}.

The annual modulation helps distinguish the Galactic DWD foreground
from an isotropic background. Our fit assigns a single amplitude to
that background, so it cannot tell us how much power comes from
AGN--IMRIs and how much comes from stellar mass EMRIs.

\par%

\section{Results}
\label{sec:results}
\label{sec:context}
We now calculate
the AGN--IMRI strain for our benchmark, upper limit, and least favorable
case, then compare the predicted spectra with the stellar mass EMRI
models of BS20 to identify
spectral differences that could help separate the AGN--IMRI and EMRI backgrounds in
the LISA global fit.\par\smallskip

\begin{figure*}[t]
    \centering
    \includegraphics[width=\linewidth]{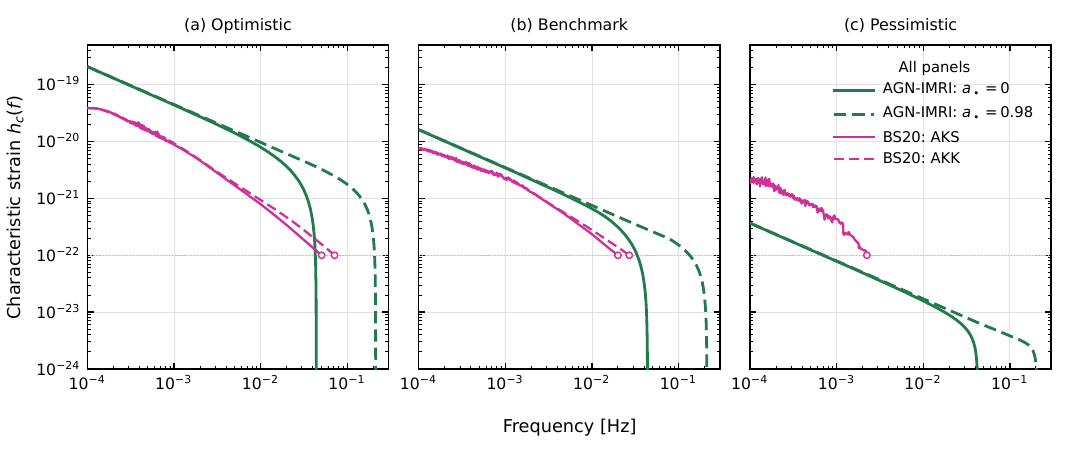}
    \caption{Prograde SMBH spin extends the AGN--IMRI background to higher
frequencies in all three panels. Green curves include all captures by
$10^5\,M_\odot$ SMBHs with $a_\bullet=0$ (solid) or $0.98$ (dashed).
Magenta curves show the unresolved stellar mass EMRI backgrounds from
Fig.~4 of BS20, after removal of sources with
four-year $\mathrm{SNR}>20$. The analytic kludge Schwarzschild (AKS, solid)
and analytic kludge Kerr (AKK, dashed) waveforms end at the Schwarzschild
and Kerr ISCOs, respectively. Open circles mark
where the published BS20 curves end at $h_c=10^{-22}$.
From left to right, our upper limit, benchmark, and least favorable
cases (13, 10, and 1~Gyr) are compared with BS20 models M12, M1, and
M11. The solid green curve and the power-law benchmark both assume zero
spin. At $3\,{\rm mHz}$, the solid green curve
in panel~(b) gives $h_c=1.62\times10^{-21}$, about $20\%$ above the
power-law benchmark of $1.35\times10^{-21}$ in
Fig.~\ref{fig:imbh-smbh-emri}. In panel~(b), the AGN--IMRI strain at $3\,{\rm mHz}$ is about twice
the unresolved stellar mass EMRI strain for either SMBH spin in our
model, motivating a joint LISA fit.}
    \label{fig:agn-imri-spin}
\end{figure*}

We evaluate Eq.~\eqref{eq:captured_mass} using the GWTC-5.0 Default-BBH rate $R_{\rm BBH}^{\rm obs}=30.4^{+6.5}_{-5.4}\,\text{Gpc}^{-3}\,\text{yr}^{-1}$ at $z=0.2$~\cite{GWTC5pop} and the benchmark parameters in Table~\ref{tab:parameters}. The implied captured IMBH mass density is $\rho_{\rm IMR}=547.2\,M_\odot\,{\rm Mpc}^{-3}$.
For a mean captured IMBH mass of $10^3\,M_\odot$, the average AGN--IMRI capture rate over 10~Gyr is
\begin{equation}
\mathcal{R} \approx \frac{\rho_{\rm IMR}}{\langle M_{\rm IMBH} \rangle T_{\rm eff}} \sim 0.05 \, \mathrm{Gpc}^{-3}\,\mathrm{yr}^{-1}.
\label{eq:agn-imri-capture-rate}
\end{equation}
\vspace{-1pt}% Allow the following wide equation to fit the REVTeX grid.
\par%

Combining Eqs.~\eqref{eq:captured_mass} and~\eqref{eq:AIMR_exact} gives the squared-strain scaling at $f_{\rm ref}=3\,{\rm mHz}$ below. Here $\mathcal H_{10}$ denotes $\mathcal H(f_{\rm ref})$ for a constant source-frame rate over $10\,{\rm Gyr}$, with source-frame cutoff $f_{\max}=0.04\,{\rm Hz}$:
\begin{widetext}
\begin{equation}
h_{c,\mathrm{bench}}^2\simeq(1.35\times10^{-21})^2
\left(\frac{\epsilon_{\rm gw}}{0.05}\right)
\left(\frac{\kappa_{\rm IM}}{0.3}\right)
\left(\frac{f_{\rm AGN}}{0.1}\right)
\left(\frac{R_{\rm BBH}^{\rm obs}}{30.4\,{\rm Gpc}^{-3}\,{\rm yr}^{-1}}\right)
\left(\frac{\langle M_{\rm BBH}\rangle}{60\,M_\odot}\right)
\left(\frac{T_{\rm eff}}{10\,{\rm Gyr}}\right)
\left(\frac{\mathcal H(f_{\rm ref})}{\mathcal H_{10}}\right).
\label{eq:AIMR_scaling}
\end{equation}
\end{widetext}
We obtain the strain error bars by propagating the GWTC-5.0 90\% credible interval on the BBH merger rate through $h_c\propto (R_{\rm BBH}^{\rm obs})^{1/2}$, with all other inputs fixed at their benchmark values.
Below the high-frequency turnover, the benchmark strain spectrum is approximately
\begin{equation}
h_c^{\mathrm{bench}}(f)
\simeq 1.35^{+0.14}_{-0.13} \times 10^{-21}
\left(\frac{f}{3\,{\rm mHz}}\right)^{-2/3}.
\end{equation}
The BNS background in the band shown comes from
inspirals, while oscillations of the merger remnants contribute at
kHz frequencies~\cite{TakamiRezzollaBaiotti2015,LehoucqDvorkinRezzolla2025}.
At $3\,{\rm mHz}$ in Fig.~\ref{fig:imbh-smbh-emri}, one can see that the benchmark strain from the AGN--IMRI GWB is within a factor of two of the strain predicted by BS20 for a stellar mass EMRI GWB.
For this first estimate, we assume a constant IMBH capture rate in AGN
disks until its redshift evolution is better understood. This gives a
conservative benchmark at $3\,{\rm mHz}$ compared with the case explored
in the \hyperref[app:capture-history]{Appendix}, where captures trace the evolving
BBH merger rate used in Ref.~\cite{Mingarelli2026}.

The mass budget that sets the benchmark strain also lets us estimate
how much these captures contribute to SMBH growth.
We adopt the local SMBH mass density $\rho_\bullet=1.8^{+0.8}_{-0.5}\times10^6\,M_\odot\,{\rm Mpc}^{-3}$ from Liepold and Ma~\cite{Liepold2024,LiepoldMa2026Erratum}, as in our companion paper. For the benchmark captured mass density, this gives
\begin{equation}
f_{\rm IMR}\simeq3.04^{+1.17}_{-0.94}\times10^{-4}.
\label{eq:fIMR_def}
\end{equation}
The benchmark captures thus deliver mass equivalent to about $0.03\%$ of the local SMBH mass density. The uncertainty in $f_{\rm IMR}$ reflects the uncertainty in the local SMBH mass density alone.

To compute the AGN--IMRI GWB upper limit, we raise all five factors in Eq.~\eqref{eq:captured_mass} from their benchmark values to the upper values in Table~\ref{tab:parameters}, giving $\rho_{\rm IMR}=9.63\times10^4\,M_\odot\,{\rm Mpc}^{-3}$. For the upper limit, we use $R_{\rm BBH}^{\rm obs}=49.4\,{\rm Gpc}^{-3}\,{\rm yr}^{-1}$, the upper end of the 90\% credible interval for the PixelPop rate at $z=0.2$ in Table~2 of Ref.~\cite{GWTC5pop}.

Even in this optimistic case, Eq.~\eqref{eq:captured_mass_fraction} gives $f_{\rm IMR}=5.35\%$ for the central value of $\rho_\bullet$. Holding $\epsilon_{\rm gw}=0.05$ and $f_{\max}=0.04$~Hz fixed, we write Eq.~\eqref{eq:AIMR_scaling} with the upper values as
\begin{widetext}
\begin{equation}
h_{c,\mathrm{UL}}^2\simeq(1.73\times10^{-20})^2
\left(\frac{\epsilon_{\rm gw}}{0.05}\right)
\left(\frac{\kappa_{\rm IM}}{1}\right)
\left(\frac{f_{\rm AGN}}{1}\right)
\left(\frac{R_{\rm BBH}^{\rm obs}}{49.4\,{\rm Gpc}^{-3}\,{\rm yr}^{-1}}\right)
\left(\frac{\langle M_{\rm BBH}\rangle}{150\,M_\odot}\right)
\left(\frac{T_{\rm eff}}{13\,{\rm Gyr}}\right)
\left(\frac{\mathcal H(f_{\rm ref})}{\mathcal H_{13}}\right),
\label{eq:AIMR_scaling_upper}
\end{equation}
\end{widetext}
where $\mathcal H_{13}$ is $\mathcal H(f_{\rm ref})$ for the 13~Gyr constant-rate history.

For the upper limit, we independently set each of the five inputs to
the upper end of its range. Forward models of black-hole growth and
mergers in AGN disks~\cite{FordMcKernan2025,Tagawa2020,Leigh2018}
could test whether a single population can realize all five values
together.

We compare the optimistic (M12), fiducial (M1), and pessimistic (M11) stellar mass EMRI backgrounds of BS20 with our upper-limit, benchmark, and least favorable inputs (Fig.~\ref{fig:agn-imri-spin}). BS20 compute these backgrounds using the EMRI population models of
Babak et al.~\cite{Babak2017}, which span assumptions about the SMBH
population and EMRI formation (e.g., the SMBH mass function and the
number of direct plunges per EMRI). Within each BS20 panel, the two curves use the same spins for the central SMBHs and differ in where the waveform is terminated. In the BS20 models M12 and M1, the central SMBHs have high spins.
In M11, the central SMBHs have zero spin, so the two waveform cutoffs
coincide. For the least favorable AGN--IMRI case, we set each of the five factors
in Eq.~\eqref{eq:captured_mass} to the lower end of its range in
Table~\ref{tab:parameters}, giving
$\rho_{\rm IMR}=0.25\,M_\odot\,{\rm Mpc}^{-3}$.

Prograde SMBH spin extends the GWB to higher frequencies by allowing the IMBH to spiral closer to the central SMBH before plunge. See Eq.~\eqref{eq:fISCO}.
For our benchmark capture history and $10^5\,M_\odot$ SMBHs, increasing $a_\bullet$ from 0 to 0.98 raises $h_c$ by only $3\%$ at $3\,{\rm mHz}$ but by a factor of $2.6$ at $30\,{\rm mHz}$, as shown in Fig.~\ref{fig:agn-imri-spin}(b).
Interpreting the cutoff as a probe of the least massive SMBHs therefore requires accounting for spin.
\par%

\section{Discussion}
\label{sec:discussion}
\label{sec:emri_distinction}
\nopagebreak[4]%

At $3\,{\rm mHz}$, the strain from all AGN--IMRI captures in our
benchmark model is within a factor of two of the strain from unresolved
stellar mass EMRIs in the BS20 M10 model (see Fig.~\ref{fig:imbh-smbh-emri}).
In our benchmark, IMBH captures accounting for $0.03\%$ of the local
SMBH mass density (see Eq.~\eqref{eq:fIMR_def}) produce this strain. The comparable strains at
$3\,{\rm mHz}$ motivate modeling both inspiral populations in the LISA
global fit to determine how much each contributes to the unresolved
background.

The AGN--IMRI background could tell us how much of the mass processed
in BBH mergers in AGN disks ultimately reaches central SMBHs as IMBHs.
In our model, a measurement or upper limit on the background amplitude,
combined with the LVK merger rate and an assumed mean BBH mass in AGN
disks, would constrain the product $f_{\rm AGN}\kappa_{\rm IM}$.
An independent estimate of the fraction of BBH mergers in AGN disks
would then let us test how efficiently their remnants reach the
central SMBH.
At migration traps in AGN disks, black holes can accumulate and grow
into IMBHs through repeated mergers~\cite{Bellovary2016,Grishin2024}.
Such disks could host frequent BBH mergers yet contribute little to
the AGN--IMRI background if most of the resulting IMBH mass never
reaches the central SMBH.

Black holes in AGN disks can also grow by accreting gas even without a
prior BBH merger. The LVK merger rate alone cannot tell us how much
mass they gain this way. Outflows can suppress accretion~\cite{Chen2023},
and in the model of \citet{Tagawa2026}, a change in the accretion state
after a merger limits the remnant's growth. Including gas accretion in
our estimate of $\rho_{\rm IMR}$ would require adding the extra mass
that captured IMBHs carry into central SMBHs.

IMBHs delivered by infalling clusters can approach central SMBHs on
eccentric orbits~\cite{ArcaSeddaGualandris2018}, while gas damping
motivates the quasi-circular inspirals in our AGN disk model~\cite{McKernan2012}.
If cluster inspirals retain appreciable eccentricity in the LISA band,
GW power in higher harmonics could help distinguish their contribution
to the background~\cite{PetersMathews1963}. Testing this possibility
requires a model for the cluster capture rate and the distributions
of IMBH masses and orbital eccentricities.
\par%

The spectrum could also tell us about the central SMBHs themselves.
In the AGN disk channel, the high-frequency cutoff is sensitive to
the least massive SMBHs that capture IMBHs. Resolved stellar
mass EMRIs measure individual SMBH masses~\cite{GairTangVolonteri2010,GairSesanaBertiVolonteri2011},
and the cutoff of their unresolved background is also sensitive to the
least massive SMBHs in that population (see BS20). Comparing the two
cutoffs could reveal whether both channels extend to similarly low SMBH
masses. Relating the AGN--IMRI cutoff to the overall SMBH mass function
requires modeling SMBH spins, AGN activity, and how the capture rate and
the masses of captured IMBHs vary with SMBH mass and redshift~\cite{Sigl2007,TucciVolonteri2017}.

Figure~\ref{fig:agn-imri-spin} illustrates how SMBH spin changes the
high-frequency AGN--IMRI spectrum at fixed SMBH mass. Eccentric stellar
mass EMRIs can also radiate at high frequencies through higher orbital
harmonics (see BS20). To test whether these spectral differences persist
in the unresolved backgrounds, we need to model the SMBH masses, spins,
and inspiral orbits in both populations and apply the same criterion
for subtracting resolved sources~\cite{Pozzoli2023,Pozzoli2024Erratum}. Electromagnetic surveys and LVK
observations could help constrain these populations before
LISA~\cite{Burke2025,ZhuChen2025}.

Even where their spectra overlap, the two backgrounds could differ
in how many unresolved inspirals contribute to a frequency bin and
how much power each supplies.
If a few sources dominate a frequency bin, their combined strain can
depart from a Gaussian distribution~\cite{Piarulli2025}. Whether the
signal is intermittent depends on the rate at which inspirals enter the
bin and how long they spend there~\cite{Regimbau2011}. A few bright inspirals can also
concentrate power in particular sky directions, even for a population
that is isotropic on average.

To test whether these spectral and statistical differences help LISA
distinguish the two backgrounds, we could simulate both inspiral
populations together with the annually modulated Galactic DWD
foreground and instrumental noise. We would then subtract the
inspirals that LISA resolves individually~\cite{GairAntonelliBarbieri2023,ChapmanBird2023,Fan2024TianQin,Singh2026EMRIPopulation}.
Comparing the amplitudes recovered by a joint fit with their input
values would show how accurately LISA could separate the AGN--IMRI and
stellar mass EMRI contributions to the unresolved background.
\par%

\section{Conclusion}
\label{sec:conclusion}

If AGN disks grow IMBHs that subsequently inspiral into their central SMBHs, the process must generate a stochastic GWB whose amplitude is set by the captured IMBH mass, its source history, and the GW efficiency. We have applied the energy relation of Ref \cite{Mingarelli2026} to a new AGN channel. The mass--budget framework gives a benchmark amplitude and an upper limit for this GWB. We find a fiducial level $h_c^{\mathrm{bench}}(3\,{\rm mHz})=1.35^{+0.14}_{-0.13}\times10^{-21}$ and a theoretical upper limit $h_c^{\mathrm{UL}}(3\,{\rm mHz})=1.73\times10^{-20}$. In this case, the captured IMBHs account for $5.35\%$ of the adopted local SMBH mass density.

The predicted amplitude motivates joint modeling with other backgrounds in LISA, while the spin-dependent extension shown in Fig.~\ref{fig:agn-imri-spin} motivates searches with decihertz detectors such as BBO and DECIGO.
In our model, a detection of the AGN--IMRI background would set a
lower bound on the fraction of BBH mergers that occur in AGN disks.
Even if every remnant reached the central SMBH, enough mergers must
occur in disks to supply the inferred IMBH mass. An independent
estimate of that merger fraction would turn the LISA measurement
or upper limit into a test of how efficiently AGN disks retain
merger remnants, grow IMBHs, and deliver them to central SMBHs.

\begin{acknowledgments}
C.\,M.\,F.\,M. is grateful to B. Burkhart, S. E. K. Ford, B. McKernan, and V. Ozolins for useful comments. She is also grateful to E. Berti and the organizers of the ``deciHz Gravitational Wave Observations on the Moon and Beyond'' workshop~\cite{Berti2026}, which motivated her to finish this work. 
This work was supported in part by the National Science Foundation under Grant PHY--2020265, and NASA's LISA Preparatory Science program under Grant 80NSSC24K0440.
\end{acknowledgments}

\section*{AI Use Disclosure}

I used OpenAI's ChatGPT (GPT-6.1 Astra, GPT-6.2 Astra, and GPT-6.2 Sol)
and Anthropic's Claude (Fable 5.1, Fable 5.2, and Opus 5.5) to assist with
literature surveys, copyediting, figure generation, and writing Python
code for the calculations. I used each tool to cross-check the other's output. A test
harness checked that newly developed code reproduced existing results.
I reviewed the manuscript and figures and take responsibility for the
final work.

\appendix*
\renewcommand{\theequation}{A\arabic{equation}}
\section{Impact of a $z$-dependent capture rate}
\label{app:capture-history}

Assume negligible delay between stellar mass BBH mergers and IMBH capture.
Following Appendix~B of Ref.~\cite{Mingarelli2026}, we use the GWTC-5.0 merger rate~\cite{GWTC5pop}
\begin{equation}
R_{\rm BBH}(z)=R_{\rm BBH}^{\rm obs}\left(\frac{1+z}{1.2}\right)^{2.5},
\end{equation}
with the same $R_{\rm BBH}^{\rm obs}=R_{\rm BBH}(0.2)=30.4\,{\rm Gpc}^{-3}\,{\rm yr}^{-1}$ as in Table~\ref{tab:parameters}.

\begin{table}[!ht]
\caption{Effect of the BBH merger history on the 10~Gyr benchmark. Spectral quantities are evaluated at $3\,{\rm mHz}$. The last row compares the strain from all AGN--IMRI captures with the unresolved stellar mass EMRI strain in model M10 of BS20 (see Fig.~\ref{fig:imbh-smbh-emri}).}
\label{tab:capture-history}
\centering
\setlength{\tabcolsep}{3pt}
\begin{tabular}{@{}lrr@{}}
\toprule
Quantity & Constant rate & Evolving rate \\
\midrule
$\overline R_{\rm BBH}$ ($\mathrm{Gpc}^{-3}\,{\rm yr}^{-1}$) & $30.4$ & $69.26$ \\
$\rho_{\rm IMR}$ ($M_\odot\,{\rm Mpc}^{-3}$), Eq.~\eqref{eq:captured_mass} & $547.2$ & $1{,}247$ \\
$\mathcal R$ ($\mathrm{Gpc}^{-3}\,{\rm yr}^{-1}$), Eq.~\eqref{eq:agn-imri-capture-rate} & $0.055$ & $0.125$ \\
$f_{\rm IMR}$, Eq.~\eqref{eq:fIMR_def} & $0.030\%$ & $0.069\%$ \\
$\mathcal H(f_{\rm ref})$, Eq.~\eqref{eq:AIMR_exact} & $0.104$ & $0.097$ \\
$\Omega_{\rm gw}(f_{\rm ref})$, Eq.~\eqref{eq:energy_distribution} & $2.25\times10^{-11}$ & $4.81\times10^{-11}$ \\
$h_c(f_{\rm ref})$, Eqs.~\eqref{eq:AIMR_scaling} and~\eqref{eq:AIMR_scaling_evolving} & $1.35\times10^{-21}$ & $1.97\times10^{-21}$ \\
$h_c(f_{\rm ref})/h_{c,{\rm M10}}(f_{\rm ref})$ & $\simeq1.6$ & $\simeq2.3$ \\
\bottomrule
\end{tabular}
\end{table}

For fixed AGN merger fraction $f_{\rm AGN}=0.1$, capture efficiency $\kappa_{\rm IM}=0.3$, and mean binary mass $\langle M_{\rm BBH}\rangle=60\,M_\odot$,
\begin{equation}
\frac{{\rm d}\rho_{\rm IMR}}{{\rm d}z}
 = \frac{\kappa_{\rm IM} f_{\rm AGN} \langle M_{\rm BBH}\rangle R_{\rm BBH}(z)}
 {(1+z)H(z)} ,
\end{equation}
where $H(z)=H_0\sqrt{0.3(1+z)^3+0.7}$ and $H_0=67.4\,{\rm km\,s^{-1}\,Mpc^{-1}}$.
Over $T_{\rm eff}=10\,{\rm Gyr}$ ($z_{\max}\simeq1.66$), the captured IMBH mass density is
\begin{equation}
\rho_{\rm IMR}
 = \int_0^{z_{\max}} \frac{{\rm d}\rho_{\rm IMR}}{{\rm d}z}\,{\rm d}z
 \simeq 1{,}247\,M_\odot\,{\rm Mpc}^{-3}.
\end{equation}
We use $\langle M_{\rm IMBH}\rangle=10^3\,M_\odot$ for the mean capture rate $\mathcal R$ and $\rho_\bullet=1.8\times10^6\,M_\odot\,{\rm Mpc}^{-3}$ for the captured mass fraction $f_{\rm IMR}$. See Eqs.~\eqref{eq:agn-imri-capture-rate} and~\eqref{eq:captured_mass_fraction}.

At $f_{\rm ref}=3\,{\rm mHz}$, $(1+z_{\max})f_{\rm ref}<f_{\max}=0.04\,{\rm Hz}$, so the full history contributes to the circular inspiral spectrum,
\begin{equation}
\begin{aligned}
\mathcal H(f_{\rm ref}) &=
    \frac{2}{3}\left(\frac{f_{\rm ref}}{f_{\max}}\right)^{2/3}
    \frac{1}{\rho_{\rm IMR}} \\
 &\quad\times\int_0^{z_{\max}} \frac{{\rm d}\rho_{\rm IMR}}{{\rm d}z}
    \frac{{\rm d}z}{(1+z)^{1/3}}
    \simeq 0.097 .
\end{aligned}
\end{equation}
The factor $(1+z)^{-1/3}$ reduces $\mathcal H(f_{\rm ref})$ when a larger share of the captured mass comes from higher redshift.
In Eqs.~\eqref{eq:AIMR_scaling} and~\eqref{eq:AIMR_scaling_upper},
the constant BBH rate is replaced by its source-time mean,
\[
\overline R_{\rm BBH}\equiv\frac{1}{T_{\rm eff}}\int_0^{z_{\max}}
       \frac{R_{\rm BBH}(z)}{(1+z)H(z)}\,{\rm d}z .
\]
Over 10~Gyr, $\overline R_{\rm BBH}=2.278\,R_{\rm BBH}^{\rm obs}$ for the evolving rate.
At $f_{\rm ref}=3\,{\rm mHz}$, Eq.~\eqref{eq:AIMR_exact} then gives
\onecolumngrid
\vspace*{\baselineskip}
\begin{equation}
h_{c,\mathrm{evol}}^2\simeq(1.97\times10^{-21})^2
\left(\frac{\epsilon_{\rm gw}}{0.05}\right)
\left(\frac{\kappa_{\rm IM}}{0.3}\right)
\left(\frac{f_{\rm AGN}}{0.1}\right)
\left(\frac{\overline R_{\rm BBH}}{69.26\,{\rm Gpc}^{-3}\,{\rm yr}^{-1}}\right)
\left(\frac{\langle M_{\rm BBH}\rangle}{60\,M_\odot}\right)
\left(\frac{T_{\rm eff}}{10\,{\rm Gyr}}\right)
\left(\frac{\mathcal H(f_{\rm ref})}{0.0973}\right).
\label{eq:AIMR_scaling_evolving}
\end{equation}
\twocolumngrid
At fixed $R_{\rm BBH}(0.2)$, the evolving rate increases the captured mass over 10~Gyr by a factor of $2.28$ (Table~\ref{tab:capture-history}).
The $1.73\times10^{-20}$ upper limit in Eq.~\eqref{eq:AIMR_scaling_upper} uses a constant rate over 13~Gyr and the upper inputs in Table~\ref{tab:parameters}.
If IMBH captures trace the evolving BBH merger rate, the higher past rate raises the $3\,{\rm mHz}$ strain by a factor of $1.46$, while the captured mass still amounts to only $0.07\%$ of the local SMBH mass density.
\par%

% Restore the journal defaults for the bibliography.
%\makeatletter
%\flushbottom
%\@booleanfalse\raggedcolumn@sw
%\makeatother

\bibliography{bib}
\end{document}